\documentclass{PoS}

\usepackage{graphicx}
\usepackage{url} 
\usepackage{wrapfig}
\usepackage{subfigure}
\usepackage{amsmath}
\usepackage{amssymb}
\usepackage{appendix}
\usepackage{alltt}
\usepackage{longtable}
\usepackage{tabularx}
\ifpdf\else\usepackage{graphics}\fi

\makeatletter
\renewcommand{\fnum@figure}{{\bf Figure \thefigure}}
\renewcommand{\fnum@table}{{\bf Table \thetable}}
\makeatother
\let\ocaption\caption
\renewcommand{\caption}[2][]{\ocaption[#1]{{\small\it #2}}}
\title{A LOFAR RFI detection pipeline and its first results}

\ShortTitle{A LOFAR RFI pipeline}

\author{\speaker{A.R. Offringa}\\
        Kapteyn Astronomical Institute, University of Groningen, The Netherlands\\
        E-mail: \email{offringa@astro.rug.nl}}

\author{A.G. de Bruyn,\\
        Kapteyn Astronomical Institute and ASTRON, The Netherlands\\
        E-mail: \email{ger@astron.nl}}

\author{S.~Zaroubi,\\
        Kapteyn Astronomical Institute, University of Groningen, The Netherlands\\
        E-mail: \email{saleem@astro.rug.nl}}

\author{M.~Biehl\\
        Institute for Mathematics and Computing Science, University of Groningen, The Netherlands\\
        E-mail: \email{m.biehl@rug.nl}}

\abstract{Radio astronomy is entering a new era with new and future radio observatories such as the Low Frequency Array and the Square Kilometer Array. We describe in detail an automated flagging pipeline and evaluate its performance. With only a fraction of the computational cost of correlation and its use of the previously introduced SumThreshold method, it is found to be both fast and unrivalled in its high accuracy. The LOFAR radio environment is analysed with the help of this pipeline. The high time and spectral resolution of LOFAR have resulted in an observatory where only a few percent of the data is lost due to RFI.
}

\FullConference{RFI mitigation workshop - RFI2010,\\
		March 29-31, 2010\\
		Groningen, the Netherlands}

\begin{document}
\label{firstpage}
\maketitle

\section{Introduction}
% Introduction in context
Now that radio telescopes of the next generation, such as the Low Frequency Array (LOFAR), the Murchison Widefield Array (MWA) and the Square Kilometer Array (SKA), are coming into operation, the dawn of software-driven telescopes producing terabyte sized data sets has begun. Because of that, data reduction in radio astronomy is entering a new era in which more emphasis is put on automated data processing and pipelining the various steps in the data reduction. One important step in the reduction process is dealing with radio frequency interference (RFI). Although trivial techniques performed by the data reducing scientist, such as manual baseline, time and frequency selection, have been sufficient albeit tedious for the last decades, more sophisticated automated flagging procedures are required for the next generation telescopes.

% Stages of RFI : Need for post-correlation RFI flagging
RFI mitigation can be performed pre- as well as post-correlation. Pre- and post-correlation techniques are mostly complimentary: they find or remove different kinds of RFI. Hence, the implementation of one does not make the other obsolete. The pre-correlation mitigation stage can detect RFI bursts of a sub-integration time changing nature with minimal loss of data, though has to be executed \emph{on-line}, implying the requirement of real-time execution on small domains, in particular on a small time interval with respect to the entire observation. Examples of pre-correlation methods are based on thresholding \cite{chi-square-time-blanking-weber, multichannel-rfi-mitigation, wsrt-rfims, pulse-blanking}; spatial filtering with eigenvalue decomposition \cite{multichannel-rfi-mitigation,hampson-spatial-nulling-2002, ellingson-spatial-nulling-2002}; and adaptive cancellation with a reference antenna \cite{adaptive-cancellation}.

To deal with RFI, the post-correlation phase is the final resort. Demonstrated techniques include the use of an independent RFI reference signal to subtract RFI \cite{post-correlation-reference-signal}; an approach using singular value decomposition \cite{post-correlation-rfi-classification, the-gmrt-eor-experiment}; and fringe fitting \cite{fringe-fitting-rfi-mitigation}. Since RFI comes in many forms \cite{rfi-mitigation-overview-fridman-baan,interference-model-lemmon} not all contaminated samples can be recovered, despite the numerous existing techniques. Therefore, flagging remains an important final step \cite{post-correlation-rfi-classification, statistical-rfi-removal, winkel-data-reduction-effelsberg}.

The situation for RFI flagging strategies in modern observatories such as the Westerbork Synthesized Radio Telescope (WSRT), LOFAR and the Giant Metrewave Radio Telescope (GMRT) has changed. On one hand, the high time and frequency resolutions make accurate flagging of contaminated samples available, resulting in smaller loss of data. On the other hand, radio quiet zones are harder to achieve, and all of the above mentioned telescopes are situated in populated areas. Moreover, sensitivity requirements for telescopes are growing. For example, one of the LOFAR key science projects is the LOFAR Epoch of Reionization (EoR) project \cite{lofar-foreground,lofar-simulations}, a very ambitious project with high demands on sensitivity and noise behaviour.

Automated flaggers can be compared on accuracy, i.e., the true/false-positive ratio; the speed of the algorithm; robustness; and technical requirements that it imposes. Constructing a flagger that performs good on all aspects is challenging. In this article, we will describe a novel flagging strategy that has been designed to take this challenge, and has been implemented in the LOFAR observatory pipeline. % TODO paper structure

\section{Method} \label{method-chapter}
\begin{wrapfigure}{r}{50mm}
  \vspace{-20pt}
  \begin{center}
    \includegraphics[width=49mm]{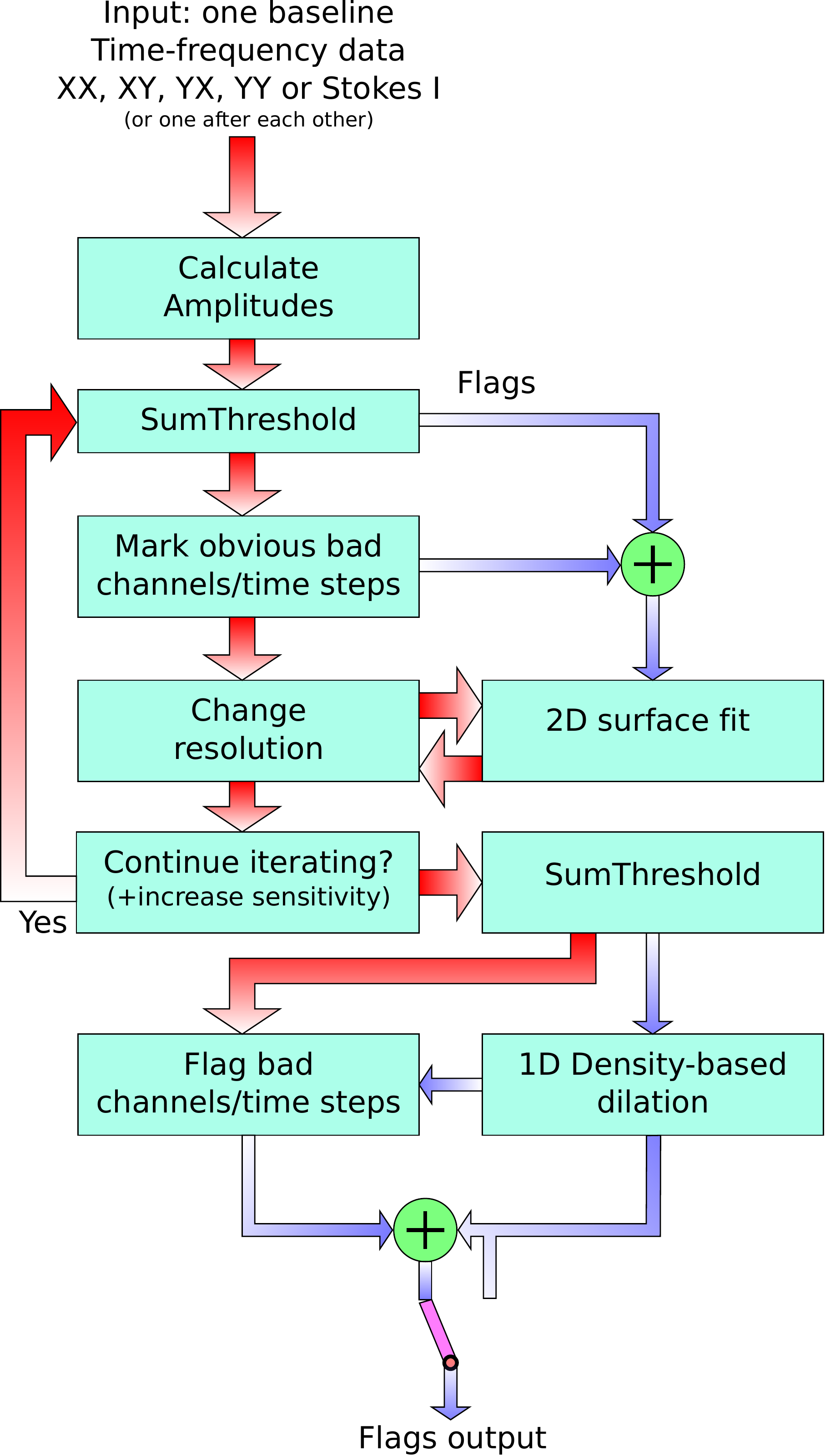}
%    \vspace{-10pt}
    \caption{Overview of the RFI flagging strategy}
    \label{fig:flagger-diagram}
  \end{center}
  \vspace{-40pt}
\end{wrapfigure}

In this section we will explain the flagger step by step. An overview of the flow of execution is given in Figure~\ref{fig:flagger-diagram}.
\subsection{Input}
The flagger is executed on the amplitude information of one polarisation of a single sub-band of a baseline. In LOFAR's common operation, a sub-band consists of 256 channels of 0.8 kHz resolution. The full band has 248 sub-bands. LOFAR can observe in two bands: the 10-80 MHz low band and the 110-240 MHz high band, which are observed by physically different antennae.

If speed is essential, the algorithm can be executed once on the Stokes-I values. Otherwise, if accuracy is more important than speed, the algorithm can be executed on the individual XX and YY or LL and RR polarisations, or on all polarisations individually. We do see some RFI that manifests in only one of the polarisations, or rotates through the polarisations, and some advantage is therefore seen when flagging all polarisations individually.

\subsection{Iterations} \label{sec-iterations}
A part of the algorithm is iterated a few times, depicted in Figure~\ref{fig:flagger-diagram} by the ``Continue iterating'' block. This is necessary for finding low-level RFI, as will be explained in the thresholding paragraph, \S\ref{sec-sumthreshold}. Iterations, however, are costly in term of speed, and should be kept to a minimum. To do so, the fit should converge quickly. We do this by entirely ignoring channels and time steps in the first surface fit that superficially look bad, yet might only have been partially uncontaminated. The extra information that might have been added if the uncontaminated part of the channel or time step was added does not change the fit much, and therefore is not slowing down the convergence.

It was determined that performing the fit two times is enough for a stable, accurate fit. This is true for all data that was tested, in special for both WSRT and LOFAR data, and for both clean bands and strongly contaminated bands.

\subsection{The \texttt{SumThreshold} method} \label{sec-sumthreshold}
The \texttt{SumThreshold} method detects series of samples with higher values than expected. Consider the orthogonal slices $\rho_d(x)$ and $\Omega_d(x)$ through the fit-subtracted visibilities $R(t, \nu)$ and the flag mask $W(t, \nu)$ respectively, where $W(t, \nu)$ evaluates to zero if the corresponding value in the mask at time $t$ and frequency $\nu$ has been flagged and one otherwise. The function parameter $x$ is scaled to have a unity step size in time or frequency, for respectively a slice through the frequency direction when $d=1$ or time direction when $d=2$. The \texttt{SumThreshold} method can now be defined by the following decision rule:
\begin{small}
\begin{eqnarray} \label{sum-threshold}
 \Omega_d^{n+1}(x) = \begin{cases}
0 & \text{if } \Omega_d^{n}(x)=0 \vee \exists i \in \{0\ldots M_n-1\}:\\
  & \hspace{1cm} \sum \limits_{j=0}^{M_n-1}
 \left|\rho_d \odot \Omega^n_d \left(x + i - j) \right)\right| > \frac{\chi_{M_n}}{\sum \limits_{j=0}^{M_n-1} \Omega^n_d \left(x + i - j\right) }\\
1 & \text{otherwise,}
\end{cases}
\end{eqnarray}
\end{small}
with parameters $\mathcal{M}=\left\{M_1, M_2, \ldots, M_{\text{max}} \right\}$, the set of combination lengths and $\chi_{M_N}$, the corresponding threshold levels for a specific combination length $M_N \in \mathcal{M}$. In the previous study \cite{post-correlation-rfi-classification}, the \texttt{SumThreshold} was introduced and was shown to produce the highest accuracy of current post-correlation RFI detection algorithms. We refer to this study for detailed information about the \texttt{SumThreshold} method.

\texttt{SumThreshold} is performed in each iteration once, before the surface fit, in order to ignore RFI when fitting. It is performed one last time when the surface fit is expected to have been converged, to establish the actual flags. To increase the stability of the strategy, the sensitivity of the \texttt{SumThreshold} method starts low, i.e., it finds only the strongest RFI, and is exponentially increased each time it is executed.

\subsection{Channel and time selection}
After \texttt{SumThreshold} has found the contaminated samples, we observe especially after the first iteration, that some channels and time scans have not been fully flagged, even though they are fully contaminated. As explained in \S\ref{sec-iterations}, this might slow down convergence, which is why a second step was implemented in order to completely, hence inaccurately but quickly, flag these channels and time steps before fitting.

In order to detect problematic channels and time steps, the values are compared based on their root mean square (RMS) values. The RMS series are Gaussian smoothed and if the difference exceeds 3.5 times the standard deviation of the sequence of differences, they are flagged completely. Optionally, this selection can be executed again as the last step in the algorithm.

\subsection{Surface fitting}
A surface fit is executed to subtract fringes caused by strong sources, in order to increase accuracy. Several fitting strategies have been tested, and all sliding window methods show similarly good results in terms of accuracy. In non-sliding window approaches such as the dimensional-independent polynomial fit described in \cite{statistical-rfi-removal}, we have observed instability near the borders of the fixed windows.

A Gaussian kernel was found to produce the best average between speed, accuracy and stability. The accuracy is not significantly different from other sliding window fitting strategies, such as a dimensional-independent polynomial fit applied on a sliding window.

Since the fit is, relative to the other operations, a time-consuming operation, the input time-frequency matrix is rescaled before fitting. The time dimension is three times reduced before fitting, and the fitted Gaussians are interpolated to restore the original scale. No significant change in accuracy was observed, which underlines that the quality of the fit is, up to some point, not a crucial aspect of accurate detection.

\subsection{Density-based dilation}
\begin{figure}
 \begin{center}
  \subfigure[Time-frequency plot without the dilation]{ \label{fig:before-dilation}
   \includegraphics[width=70mm]{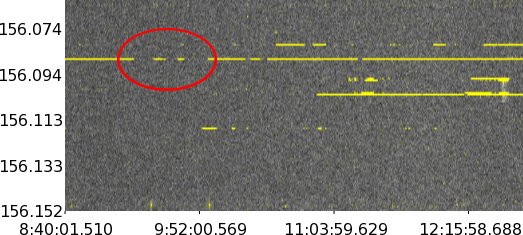}
  }
  \subfigure[Time-frequency plot after dilation]{ \label{fig:after-dilation}
   \includegraphics[width=70mm]{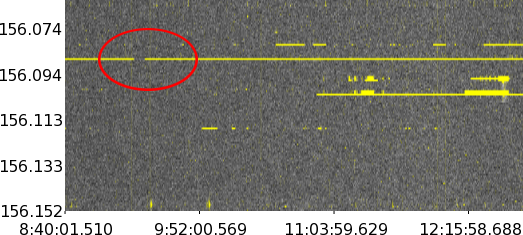}
   }
 \end{center}
 \caption{The result of a density-based dilation with $\eta = 0.9$: the flags in panel~\subref{fig:before-dilation} are established by the \texttt{SumThreshold} method and dilated based on the flag density. The result is shown in panel \subref{fig:after-dilation}. Noticeable differences are the small gaps in orthogonal lines that have been filled by the dilation, such as the area within the red ellipse. While this diagram displays over 6000 time steps, the algorithm also fills many invisible small holes: its behaviour is scale invariant.}
 \label{fig:dilation}
\end{figure}

It may be desirable to flag samples that are up to a few channels away from strong, continuous RFI. Thresholding does not flag these samples, if they are not significantly different in amplitude. Likewise, it may be desirable to flag more of a partially flagged channel, because a continuous transmitter might be recorded at different amplitudes, either because of different propagation of the signal, because of the transmitter moving in respect with the beam or because of a transmitter's intrinsically changing strength, and this might cause the received RFI not to trigger the threshold in some samples. To overcome this problem, we enlarge the flag mask after the apparent RFI has been flagged by the iterative procedure.

A typical approach in this problem is to perform a morphological dilation operation on the flag mask. For example, a dilation with a square mask of size $N\times N$ would enlarge each flag to a square of $N \times N$. Every sample, that has an orthogonal distance smaller than $N$ samples from a flagged sample, would be flagged in this case. Although this technique is advantageous for its simplicity and establishment in the field of mathematical morphology, using this technique for the described purpose has the disadvantage of being inaccurate: it will typically flag too many samples when only a few samples are flagged in some area, while too few samples will be flagged when a channel or time step is almost completely flagged.

To correct for these problems, we introduce a dilation operation which mask size is related to the one dimensional flag density: the dilation mask is larger in dense areas and smaller in sparse areas, in respect to either the one dimensional time domain or frequency domain.

Consider an orthogonal slice $\Omega_d(x)$ through the flag mask as defined in \S\ref{sec-sumthreshold}. The following density-based decision rule is introduced:
\begin{equation} \label{eq:dilation}
 \Omega_d'(x) = \begin{cases}
0 & \text{if } \exists Y_1 \le x,\ \exists Y_2>x: \sum\limits_{y = Y_1}^{Y_2-1} \Omega_d(y) \le \eta \left( Y_2 - Y_1 \right) \\
1 & \text{otherwise,}
\end{cases}
\end{equation}
where $\eta \in [0,1]$ is the density ratio threshold. In words, this rule flags the samples that are in any constructable area $\left[Y_1; Y_2\right>$ with an unflagged sample ratio less or equal than $\eta$. Specifically, $\Omega'(x)=0$ for all $x$ if $\eta=1$, while $\Omega'(x)=\Omega(x)$ for $\eta=0$. Furthermore, since any element $x$ with $\Omega(x)=0$ will be in the single element area containing only itself, $\Omega(x)=0 \implies \Omega'(x)=0$. Consequently, the number of flags is increasing. Although a strict implementation of \eqref{eq:dilation} will take $O \left( n^2 \right)$ operations for $n$ samples in the orthogonal slice $\Omega_d(x)$, by putting extra constraints on $Y_1$ and $Y_2$, an $O \left( n \log n\right)$ implementation is possible without much loss of its accuracy. Figure~\ref{fig:dilation} shows the result of such a dilation on actual data.

\section{Computational requirements}
\footnotetext[1]{Floating point operations per sample} \footnotetext[2]{These are actually integer operations, since this step uses the masks.}
Table~\ref{tbl-compreq} shows an estimate of the required floating point operations per sample for each individual step. The total number of operations required is on the order of 300 floating point operations per sample. In a typical full LOFAR observation, the correlator will output $4$~polarisations~$\times$ $256$~channels/sub-band~$\times$ $248$~sub-bands~$\times$ $\frac{50^2}{2}$~baselines~$\times$ $1$~sample/second $\approx 0.3$ gigasamples per second, yielding a computational requirement of $\sim$0.1 TFLOP/s in the best flagging mode.
\begin{wraptable}{r}{80mm}
\vspace{-15pt}
\caption{\small{Computational requirements of the RFI pipeline}} \label{tbl-compreq}
\scalebox{0.83}{
\begin{tabular}{|l|r|r|r|}
\hline
Step & F/smp\footnotemark[1] & Count & Total F/smp\footnotemark[1] \\
\hline
 Calculating amplitudes   &   4 & 1 &   4 \\
 Time/frequency selection &   2 & 3 &   6 \\
 \texttt{SumThreshold}    &  20 & 3 &  60 \\
 Change resolution        &   4 & 2 &   8 \\
 Surface fit              &  50 & 2 & 100 \\
 Density-based dilation   & 100\footnotemark[2] & 1 & 100 \\
\hline
 \multicolumn{3}{|l|}{Total} & 278 \\
\hline
\end{tabular} } % scale box end 
%\vspace{-10pt}
\end{wraptable} 
% end of table

Although this is only a small fraction of the required computations for correlation, some simplifications can be made to lower the computational requirements. Techniques to improve the computational performance include: flagging on Stokes-I values; using a larger resizing factor before fitting; using a smaller window size; and determining the cross-correlation flag masks using autocorrelations.

The LOFAR flagging pipeline will be run on an off-line computing cluster. The flagging pipeline is parallellized by running each sub-band on a different computational node, and the flagging of the individual sub-bands is executed by a multi-threaded implementation. Concluding from the interpolation of the performance of the current implementation of the pipeline, which achieves processing 27 stations in a quarter of the observing time with its most computational expensive flagging strategy, real-time performance can be realised in a full 50 station LOFAR.

\section{Input/output requirements} \label{sec-iorequirements}
Processing baseline by baseline in a pipeline has implications for the software architecture of the observatory: since baselines are correlated simultaneously, the observed visibilities have to be written to disk before running the RFI pipeline, which is inefficient. After finishing observing, the flagging pipeline can read the data in the required order. However, flagging is normally followed by tasks such as calibration and source subtraction. These tasks expect time-sorted data, thereby requiring a second read of the data in its previously observed order. Since the architecture of LOFAR allows this flow of processing, and because of the advantages of baseline by baseline flagging in terms of accuracy and computational speed, the input/output-overhead caused by this deficiency is ignorable. This however might become a serious issue in even larger telescopes such as the SKA.

\section{Flagging results} \label{sec-results-chapter}
\begin{figure}
 \begin{center}
  \subfigure[Time-frequency plot before flagging]{ \label{fig:sb154-tf-before}
   \includegraphics[width=70mm]{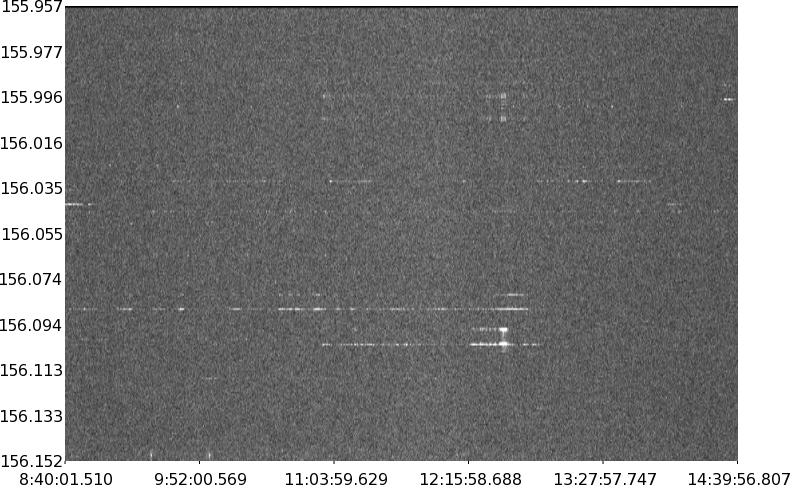}
   }
  \subfigure[Time-frequency plot after flagging]{ \label{fig:sb154-tf-after}
   \includegraphics[width=70mm]{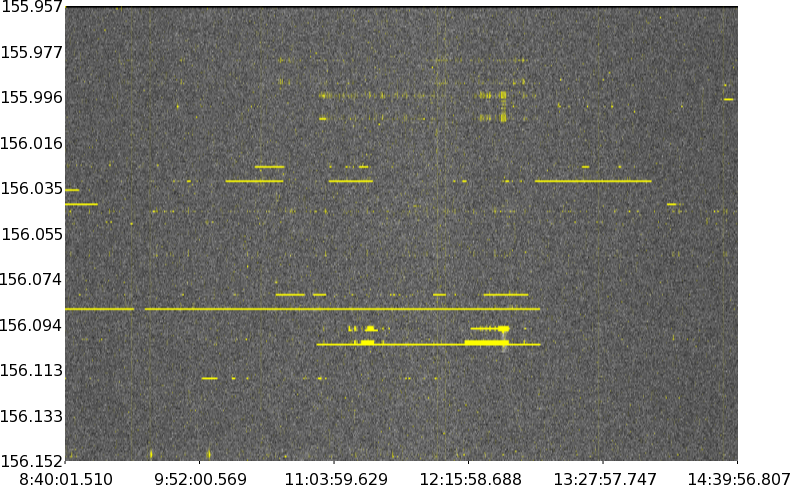}
  }
  \subfigure[Amplitude plot before flagging]{ \label{fig:sb154-scatter-before}
   \includegraphics[width=45mm]{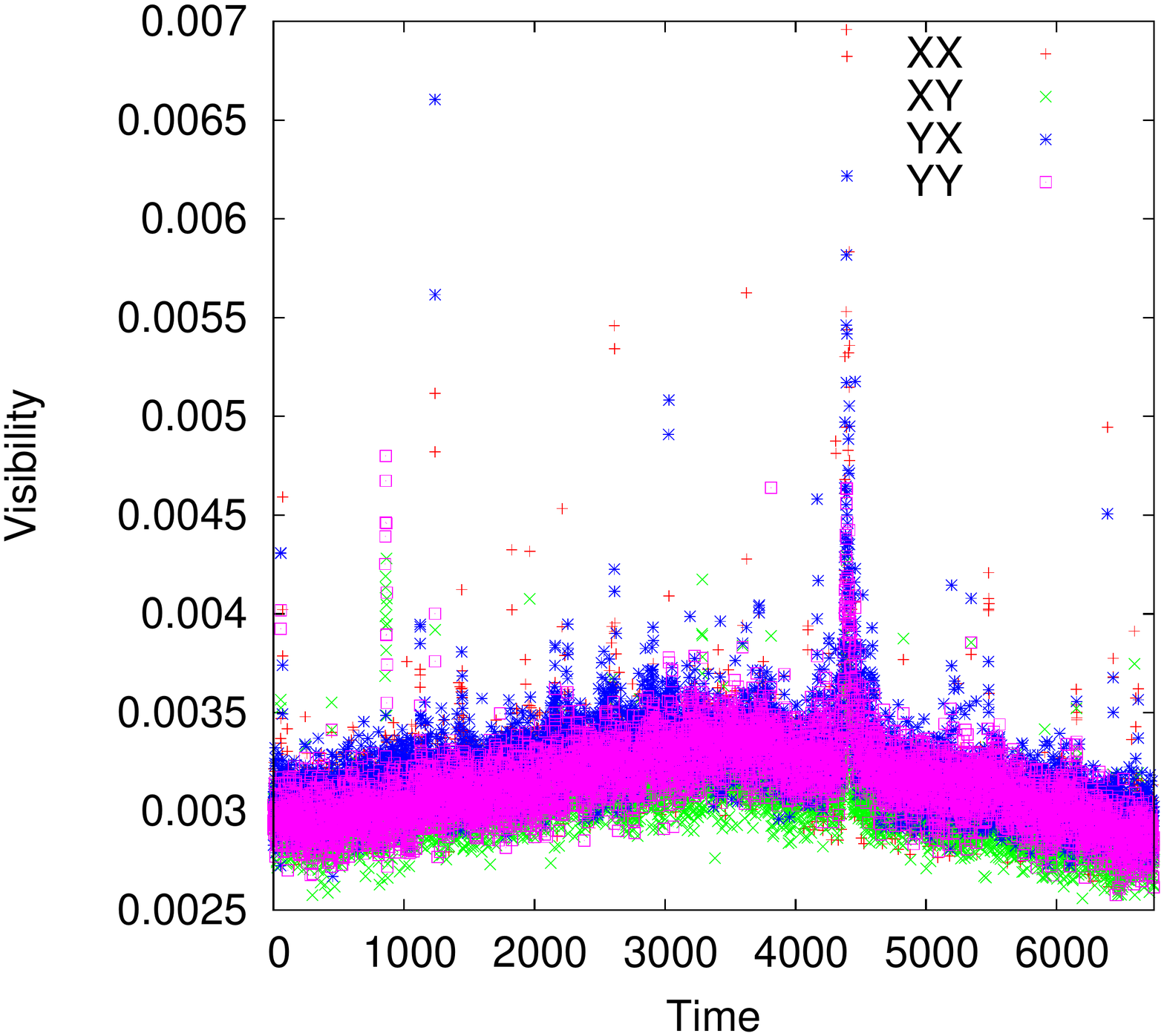}
  }
  \subfigure[Amplitude plot after flagging]{ \label{fig:sb154-scatter-after}
   \includegraphics[width=45mm]{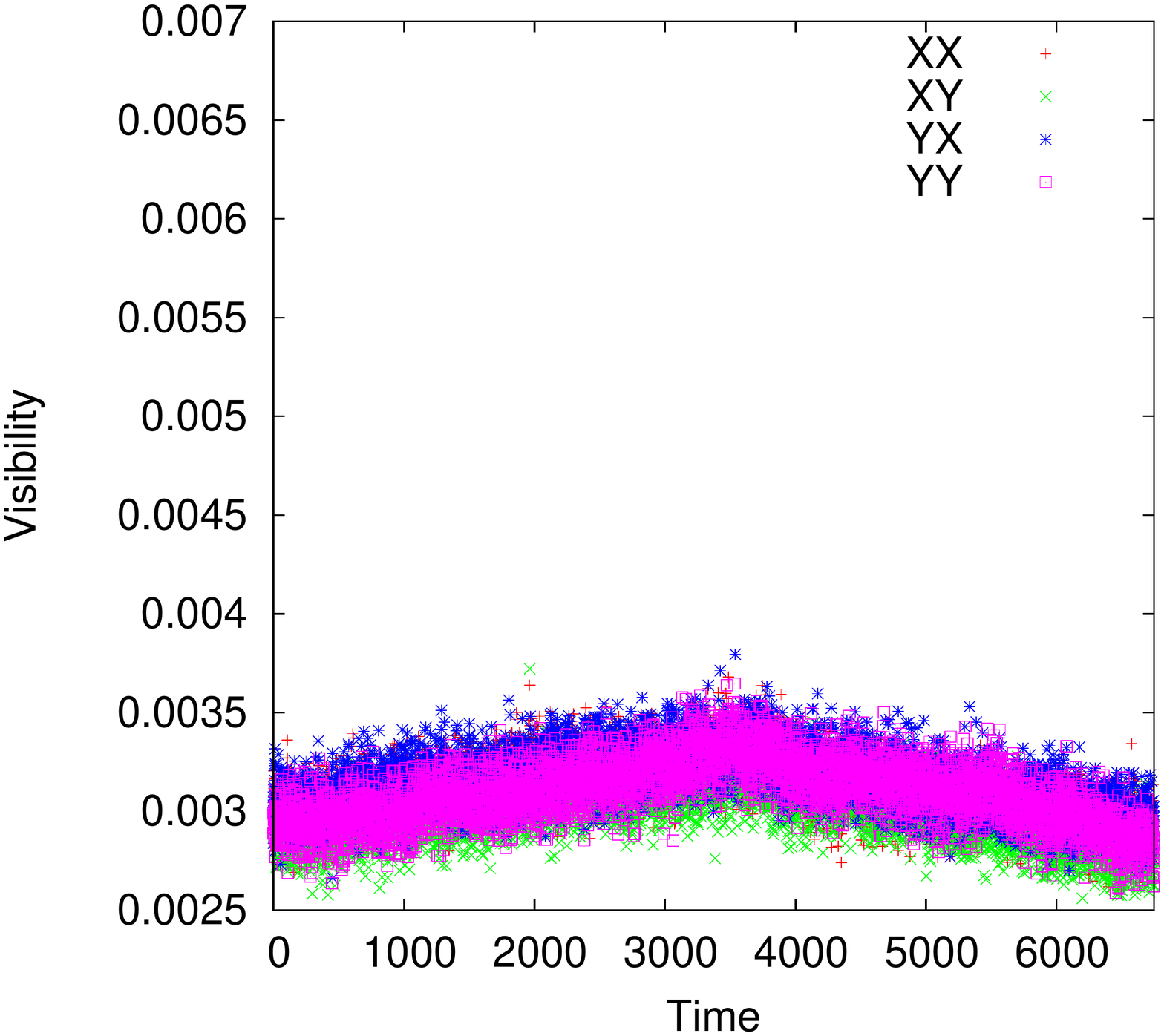}
  }
  \subfigure[Power spectrum]{ \label{fig:sb154-power-spectrum}
   \includegraphics[width=45mm]{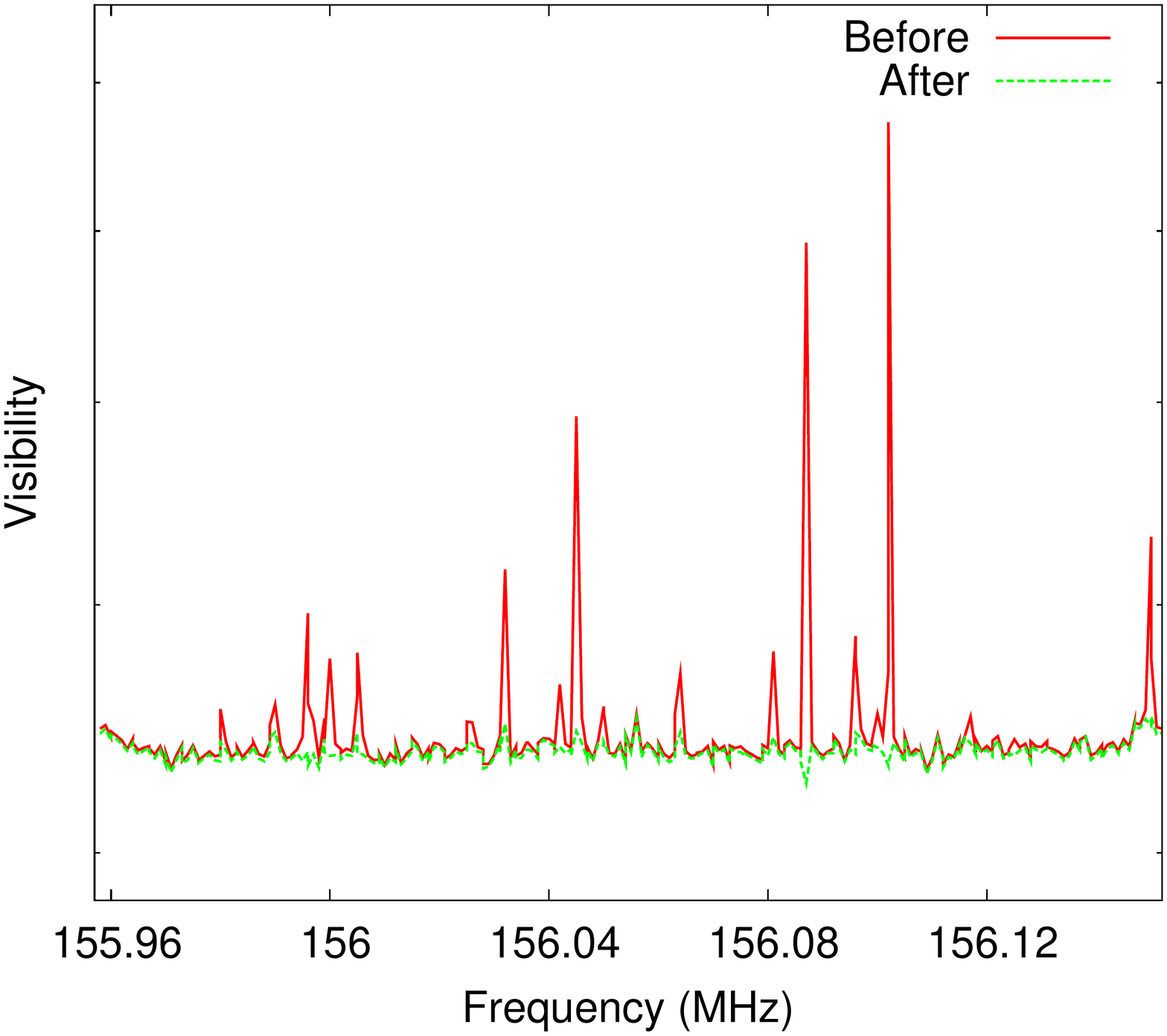}
  }
 \end{center}
 \caption{Flagging results of the 6 hour LOFAR observation L2010\_07096 of April 24, 2010. All plots show the same randomly chosen sub-band around 156 MHz for a 1.5-km baseline (CS302\_HBA1~$\times$~CS005\_HBA0) with three second integration time. The flagging pipeline was run with its default settings, and 1.8\% of the data is flagged. As can be seen from panels \subref{fig:sb154-tf-before}, \subref{fig:sb154-scatter-before} and \subref{fig:sb154-power-spectrum}, this sub-band contains relatively many interfering transmitters, yet all of them are relatively weak. The panels \subref{fig:sb154-tf-after},  \subref{fig:sb154-scatter-after} and \subref{fig:sb154-power-spectrum} show the cleaned band after flagging.}
 \label{fig:sb154}
\end{figure}
The implementation\footnote[3]{The software implementation of the presented RFI pipeline has been made publicly available and can be downloaded from the following location: \url{http://www.astro.rug.nl/rfi-software/}} of the algorithm was tested on several LOFAR observations. Currently, 27 of the approximately 50 total LOFAR stations are ready. Flagging a single sub-band of a 6 hour observation with the 27 stations takes 90 minutes on a single cluster node. This implies real-time flagging speed for the full 50 station LOFAR that will produce four times more data. All RFI that can be found by visual inspection is typically flagged, thereby outperforming simpler methods such as a median absolute threshold filter in both accuracy and speed. An example result can be found in Figure~\ref{fig:sb154}.

The pipeline is, by the use of the \texttt{SumThreshold} method, very accurate. In some cases, the algorithms finds RFI which is invisible by eye on full scale time-frequency diagrams, but becomes only apparent when zooming in on the data and integrating certain cuts of the data cube. In the band shown in figure~\ref{fig:sb154} an interferer is visible at approximately 156.03 MHz. Although it is visible as a small bump in the time integrated spectrum in Figure~\ref{fig:sb154-power-spectrum}, it is not apparent in the time frequency plot of Figure~\ref{fig:sb154-tf-before}. Nevertheless, the algorithm finds the samples that are contaminated by the interferer, and the particular bump at 156.03 MHz in Figure~\ref{fig:sb154-power-spectrum} is flattened.

On the other hand, if an interferer has a smooth time-frequency profile, it will be mistaken for astronomical data and will not be flagged. In these situations it might help to subtract a rough model for the celestial signal and increase the flagger's sensitivity.

\section{LOFAR RFI environment} \label{sec-lofar-rfi-environment}
LOFAR breaks the tradition of building telescopes in sparsely populated areas, with its core being installed in the North-East of the Netherlands. Although the core is in a nature reserve, and therefore in a sparser populated part of the Netherlands, all the stations are relatively close to farms, roads and some nearby municipalities. Now that LOFAR is half-way ready and performing representable observations, we can start to evaluate the dynamic radio environment.

The first results of RFI mitigation show several promising characteristics of the LOFAR site. First of all, hardly any broadband RFI is observed. If observed, it is typically caused by electrical fences, lightning, power cables, hardware in situ, cars and trains. It can be concluded that the site is sufficiently remote and hardware on site is sufficiently shielded to prevent these interferences. Only one of the stations is close to an electrical fence that surrounds a farming meadow, causing broadband spikes every two or three seconds. The flagging pipeline flags 40\% of the data in this station, and this station is therefore currently not useful. Options include negotiation with the farmer to switch off the electrical fence during observations or implementing an RFI nulling method in the station that nulls spikes on a high time resolution.

A second class of interferers are constant transmitters at a fixed frequency, such as FM radio. The FM range lies between the physically separated low and high bands. Transmitters in this range are therefore effectively blocked by the bandpass filters. Other constant sources that do transmit within the observing frequency often occupy only one or a few 0.8-kHz channels, which, after the pipeline has flagged these transmitters, cause only a minimal amount of data loss. While many of the sub-bands of 256-channels are completely clean of such constant transmitters, others have a few of such transmitters, such as the one shown in Figure~\ref{fig:sb154}.

A third class of interferers are transient sources with variable frequency. These occur mostly at random and their exact origin is often unknown. Some of these can be caused by moving objects, such as meteors or airplanes, that reflect a distant signal for a short period.

Considering all classes of interferers, typical observations with representable stations show only a few percent of data loss due to interference.

\section{Conclusion and discussion} \label{sec-conclusion-chapter}
Radio astronomy is entering a new era with futuristic observatories such as LOFAR and the SKA. In this article we have presented a flagging technique that has shown the ability to operate accurately and efficiently on the LOFAR observations. Therefore, this technique is also a good basis for future observatories.

Because the computational costs of the RFI pipeline are only a fraction of the correlation costs, efficiently ordering the data before presentation to an RFI algorithm is the largest challenge, rather than optimising the computational costs. The pipeline also stipulates the importance of flexibility in an observatories' architecture, which adds freedom to design decisions. The LOFAR architecture also allows more sophisticated variations that include RFI mitigation at station level and different pipelines based on the observation mode. With the example of a complicated pipeline as described in this paper, it can be concluded that other algorithms such as transient detection and other pattern recognition techniques can be implemented in a similar manner in the pipeline.

Both the software and hardware of LOFAR are still under construction. The first observations of LOFAR nevertheless show very good prospects for the telescope, with only a few percent lost data due to interferers and, highly important, neither broadband nor in situ interference is commonly seen. The next step in RFI mitigation is to produce and analyse images on a maximum dynamic range, in order to analyse the effects of possible weak RFI that is undetectable in post-correlated time frequency domains. Prevention of new transmitters remains very important, and establishment of a radio-quiet zone, especially around the core, is recommended.

In order to improve data quality further, pre-correlation techniques might be added at station level or during correlation. An interesting improvement to the robustness of a correlator might be to execute the \texttt{SumThreshold} method prior to correlation. Considering the accuracy gain of the \texttt{SumThreshold} compared to normal thresholding, and considering the correspondence of RFI on small and large timescales, implementing this pre-correlation method on the highest time resolution data might improve blanking accuracy further.

\acknowledgments

We thank the LOFAR collaboration for providing the data on which the flagger was developed and tested.

\bibliographystyle{plain}
\bibliography{LOFAR-RFI-pipeline-Offringa}

\label{lastpage}

\end{document}